\documentclass[twocolumn,pra,aps,showpacs]{revtex4-1}

\usepackage{url,psfrag,graphicx}
\usepackage{dcolumn}
\usepackage{amsmath,amssymb}
\usepackage{bm}
\usepackage{pstricks}
\usepackage{hyperref}
\usepackage{epsfig}

\def\<{\left<}
\def\>{\right>}
\def\ket|#1>{\left|#1\right>}
\def\bra<#1|{\left<#1\right|}
\def\elem<#1|#2|#3>{\left<#1\right|#2\left|#3\right>}
\def\({\left(}
\def\){\right)}

\begin{document}

\title[Short Title]{Physical consequences of P$\neq$NP and the
  DMRG-annealing conjecture}

\author{Javier Rodr\'{\i}guez-Laguna} 
\affiliation{Mathematics Dept. \& GISC, Universidad Carlos III de Madrid, Spain}

\author{Silvia N. Santalla}
\affiliation{Physics Dept. \& GISC, Universidad Carlos III de Madrid, Spain}

\date{April 22, 2014}

\begin{abstract}
Computational complexity theory contains a corpus of theorems and
conjectures regarding the time a Turing machine will need to solve
certain types of problems as a function of the input size. Nature {\em
  need not} be a Turing machine and, thus, these theorems do not apply
directly to it. But {\em classical simulations} of physical processes
are programs running on Turing machines and, as such, are subject to
them. In this work, computational complexity theory is applied to
classical simulations of systems performing an adiabatic quantum
computation (AQC), based on an annealed extension of the density
matrix renormalization group (DMRG). We conjecture that the
computational time required for those classical simulations is
controlled solely by the {\em maximal entanglement} found during the
process. Thus, lower bounds on the growth of entanglement with the
system size can be provided. In some cases, quantum phase transitions
can be predicted to take place in certain inhomogeneous
systems. Concretely, physical conclusions are drawn from the
assumption that the complexity classes {\bf P} and {\bf NP} differ. As
a by-product, an alternative measure of entanglement is proposed
which, via Chebyshev's inequality, allows to establish strict bounds
on the required computational time. 
\end{abstract}


\maketitle


\section{\label{introduction}Introduction}

Global optimization is one of the most important computational
problems in science and technology. But beyond its practical
relevance, it is also of deep theoretical interest when viewed from
the broader perspective of computational complexity theory
\cite{Papadimitriou_94,Complexity_10}. Problems are ranged into an
intrincate classification by theoretical computer scientists, and an
impressive corpus of theorems and conjectures has been built to relate
them, such as the Cook-Levin theorem \cite{Cook_71,Trakhtenbrot_84}
which proves the existence of {\bf NP}-complete problems, or the
conjecture that {\bf P}$\neq${\bf NP}.

All those complexity classes are defined with respect to an abstract
computer, the {\em Turing machine}. Physical devices designed to solve
a particular problem need not be subject to that restriction, i.e.: an
{\bf NP}-complete problem {\em might} be solved in polynomial time by
a physical device even if {\bf P}$\neq${\bf NP}. The reason is that
{\em Nature need not be a Turing machine}. Notwithstanding, {\em
  simulations} of physical processes on classical computers are bound
by the previous hierarchy of classes, since they are (approximately)
Turing machines. If {\bf P}$\neq${\bf NP}, any attempt to solve an {\bf
  NP}-complete problem in polynomial time with a simulation of a
physical process on a classical computer must fail. The reasons for
the failure must be deducible from the simulation details, and insight
about the underlying physical process might be obtained.

Quantum mechanics provides the most promising physical attempt to
outperform classical computation, and among all the quantum
computational techniques, we will focus on {\em adiabatic quantum
  computation} (AQC), also known as {\em quantum annealing}
\cite{Farhi_01,Das_08}. The possibility of using AQC to solve {\bf
  NP}-complete problems in polynomial time is one of the most exciting
problems in quantum computation, but it is not addressed in this
work. We will focus on {\em classical simulations} of AQC built upon
{\em matrix product states} (MPS)
\cite{Vidal_03,PerezGarcia_07}. Concretely, we will analyse a
technique based on an adiabatic extension of the density matrix
renormalization group (DMRG) \cite{White_93,Schollwock_11}, published
as {\em quantum wavefunction annealing} (QWA) \cite{Laguna_07}. Along
with it, we will put forward and discuss the {\em DMRG-annealing
  conjecture}, which states that the efficiency of the QWA simulations
of AQC is controlled uniquely by the maximal entanglement found during
the process.

Being a classical computational technique, DMRG-based simulations of
AQC can never solve {\bf NP}-complete problems in polynomial time,
unless {\bf P}={\bf NP}. Accepting {\bf P}$\neq${\bf NP} and the
adiabatic DMRG-conjecture to hold we can put lower bounds on the
behaviour of entanglement during AQC processes. In some cases, it
allows us to predict the existence of quantum phase transitions.

Our work, thus, puts under a different light ideas that are known in
the area. In recent years a new field is being built, known as {\em
  hamiltonian complexity}, which considers the computational
complexity of performing classical simulations of quantum systems (see
\cite{Osborne_11} for a recent review). In 2003 Vidal showed that a
digital quantum computation involving a fixed amount of entanglement
could be efficiently simulated using a matrix product representation
\cite{Vidal_03}, thus showing that an exponential speed-up was only
possible if the MPS bond dimension grows with the input size. But,
assuming that the ground state (GS) of a certain hamiltonian can be
described as a MPS of fixed dimension, how hard can it be to find it?
In 2006, Eisert showed that this problem can be NP-complete
\cite{Eisert_06}. Indeed both results are not hard to reconcile within
the DMRG-annealing conjecture framework, as we will show. In 2009
Hastings proved that AQC with fixed gap in 1D would never achieve an
exponential speed-up \cite{Hastings_09}, based on his first rigorous
proof of an area law in 1D \cite{Hastings_07}.

This paper is structured as follows. In section \ref{model} we review
the basics of adiabatic quantum computation (AQC) (or quantum
annealing) \cite{Farhi_01,Das_08}. Section \ref{qwa} details the
quantum wavefunction annealing technique, an adiabatic extension of
DMRG, which is illustrated in section \ref{sec:illustration}. The {\em
  DMRG-annealing conjecture} is formulated and discussed in section
\ref{conjecture}. Assuming this conjecture to hold, our main results,
which are the physical implications of complexity theory, are exposed
in section \ref{thesis}. The paper closes with the conclusions and
suggestions for further work.


\section{\label{model}Adiabatic quantum computation}

Since the seminal article of R.P. Feynman in 1982 \cite{Feynman_82},
physicists have had an increasing interest in the {\em simulability}
of quantum mechanics, which has grown into the field of {\em
  hamiltonian complexity} \cite{Osborne_11}. The difficulties reside
in the exponential growth of the dimension of the Hilbert
space. Quantum computation was born with the idea of converting this
handicap into an opportunity: perhaps clever exploitation of this
exponential growth will allow us to achieve an exponential speed-up of
classical algorithms, maybe even to solve {\bf NP}-complete problems
in polynomial time \cite{Nielsen_Chuang_00}. This hope has not yet
been either fulfilled or disproved, and we will not address it here.

Among the quantum computational techniques proposed, we will focus on
adiabatic quantum computation (AQC) \cite{Farhi_01}, studied also
under the name of quantum annealing
\cite{Das_08,Das_Chakrabarti_05}. AQC was proved in 2004 to be {\em
  universal} in the sense that the results of any quantum computation
can be simulated in polynomial time with an AQC \cite{Aharonov_04}.

An AQC is implemented by a physical device which establishes an
adiabatic route between two hamiltonians, $H_0$ and $H_1$, such that
the ground state (GS) of $H_0$ is easy to obtain physically, and the
GS of $H_1$ provides the solution to some problem. The GS of a
hamiltonian is difficult to achieve experimentally when the system is
subject to {\em ageing}, i.e.: when the low energy spectrum is
complex, as it happens for most disordered systems. The adiabatic
theorem ensures that, {\em if the process is slow enough and the gap
  never vanishes exactly}, the ground state of $H_1$ will be obtained
from that of $H_0$.

As a relevant example throughout this work let us consider the
(classical) spin-glass problem \cite{Binder_86}. Given a graph ${\cal
  G}$ of $N$ spins and a set of arbitrary real coupling constants
$J_{ij}$ attached to each graph link, we define the (classical)
spin-glass energy as

\begin{equation}
E=-\sum_{\<i,j\>} J_{ij} \sigma_i \sigma_j
\label{classical.sg}
\end{equation}

Where the $\sigma_i$ are values in $\{-1,+1\}$ attached to each
site. The (classical) spin-glass problem is to find the values for
$\sigma_i$ which minimize the previous energy.

If the graph is 1D, the problem is trivially in {\bf P}. If it is 2D,
a non-trivial construction found by Barahona \cite{Barahona_82} also
renders the problem polynomial. For higher dimensions, or for
arbitrary graphs of fixed connectivity ($\geq$3), the problem is {\bf
  NP}-complete \cite{Liers_03}. Even a 3D graph composed of two flat
layers yields an {\bf NP}-complete problem\cite{Barahona_82}.

The AQC strategy for the spin-glass problem sets the target
Hamiltonian, $H_1$, to be a quantum counterpart of
eq. (\ref{classical.sg}), promoting the $\{-1,+1\}$ values of
$\sigma_i$ to spin-1/2 operators \cite{Santoro_02}:

\begin{equation}
H_1=-\sum_{\<i,j\>} J_{ij} S^z_i S^z_j
\label{h0}
\end{equation}

On the other hand, the initial Hamiltonian, $H_0$, must be chosen in
such way that its GS is easy to obtain and corresponds to a system
subject to very strong quantum fluctuations. A suitable example is the
coupling to a uniform transverse magnetic field: $H_0=-\sum_i
S^x_i$. The system will interpolate adiabatically between both
Hamiltonians. At all times, it will be described by the Ising model in
a transverse field (ITF) with arbitrary couplings:

\begin{eqnarray}
H(\lambda) &=& (1-\lambda) H_0 + \lambda H_1 \\ &=& 
- \lambda \sum_{\<i,j\>} J_{ij} S^z_i S^z_j - (1-\lambda) \sum_i S^x_i 
\label{ritf}
\end{eqnarray}

\noindent where we see that $H(0)=H_0$ and $H(1)=H_1$, and $\lambda$
will be termed the {\em adiabatic parameter}.

Let $\ket|\Psi(\lambda)>$ denote the ground state of the previous
system as a function of $\lambda$, which is only degenerate for
$\lambda=1$. If $\lambda=0$, the ground state is found just by making
all spins point in the $X$-direction:

\begin{equation}
\ket|\Psi(0)>={1\over\sqrt{2}} \( \ket|+> + \ket|-> \)^{\otimes n}
\end{equation}

In this state, all classical configurations take exactly the same
probability, so we may say that it is absolutely disordered. For
$\lambda\to 1^-$, on the other hand, the ground state provides the
solution to the classical spin-glass problem. Thus, the AQC strategy
is to take $\lambda=0$, increase it adiabatically until $\lambda=1$
and then read the solution. The adiabatic theorem can be applied if
the process is slow enough, assuming that the gap never vanishes
exactly.

Of course, in the laboratory it is more convenient to leave the
coefficient of $H_1$ untouched through the procedure. The AQC strategy
is to apply a large transverse field initially, and decrease it slowly
enough \cite{Santoro_02}. The main difficulty during an AQC experiment
is to ensure adiabaticity. The probability of a jump to an excited
state increases exponentially as the energy gap closes, as reflected
by the Landau-Zener formula \cite{Santoro_02}. Thus, if the system
undergoes a quantum phase transition and the energy gap vanishes, the
velocity must be reduced in an appropriate way at that point,
increasing the computational time.

It may be tempting to try to extract conjectures about the minimal gap
along an AQC trajectory from the (classical) complexity class of the
problem at hand. But these inferences are {\em not valid}, since the
precise nature of the relation between the quantum and the classical
complexity classes is not straightforward. Recent results of Altshuler
and coworkers \cite{Altshuler_10} cast doubts on the possibility of
solving {\bf NP}-complete problems in polynomial time using quantum
computation, due to the very narrow gap distribution in disordered
systems which can be expected from Anderson's theorem. Nonetheless,
other authors are more optimistic, believing that a route which avoids
exponentially small gaps is feasible \cite{Dickson_11}. Those
problems, which are of uttermost importance for quantum computation,
will not be studied in this work.

A caveat is in order: AQC is not designed for cases in which the gap
is exactly zero at some point. This imposes certain restrictions on
$H_0$ and $H_1$. For example: an adiabatic calculation in which the
Hamiltonian interpolates linearly between the (classical)
ferromagnetic Ising model in $S^z$ and a (classical) spin-glass, also
in $S^z$, is not possible. Both Hamiltonians commute, the ground state
becomes degenerate at least once, and the adiabatic theorem does not
apply.


\section{\label{qwa}Classical simulation of AQC: quantum wavefunction
  annealing}

In order to apply the results of computational complexity theory, we
should analyse algorithms running on Turing machines, not on arbitrary
physical devices. Therefore, we will study {\em simulations} of
adiabatic quantum computation running on a classical computer.

A first simulation approach to AQC is the use of path integral
Monte-Carlo methods (PIMC) \cite{Santoro_02}. This technique does not
suffer from Landau-Zener (avoided) level crossings, and the closing of
the energy gap does not constitute a problem. Nonetheless, if an
attempt is made to solve an {\bf NP}-complete problem using it, it is
always found that, at some moment, the system undergoes {\em critical
  slowing down}. This forces long relaxation times and reduces the
efficiency of the procedure. The exact amount of this reduction is not
easy to quantify, due to the different complexity classes associated
with probabilistic computation \cite{Battaglia_05}.

A different simulation procedure, quantum wavefunction annealing
(QWA), is a fully deterministic classical algorithm and lends itself
more easily to analysis \cite{Laguna_07}. The key feature of QWA
simulation is that it computes the full wavefunction of the involved
ground states. Let $H(\lambda)=(1-\lambda)H_0 + \lambda H_1$, with
$\lambda \in [0,1]$. Then, the QWA procedure is:

\begin{enumerate}

\item Let $\lambda=0$ and find the GS of the initial hamiltonian
  $H(0)=H_0$, $\ket|\Psi(0)>$. Choose a suitable initial value for
  $\delta\lambda$.

\item Increase the adiabatic parameter: $\lambda'=\min(\lambda+\delta
  \lambda,1)$.

\item Find the GS of $H(\lambda')$, $\ket|\Psi(\lambda')>$, {\em using
  the previous ground state as a seed}.

\item If the overlap $|\left<\Psi(\lambda')|\Psi(\lambda)\right>|$ is
  below a given threshold, halve $\delta \lambda$ and return to 2.

\item If $\lambda'<1$, make $\lambda=\lambda'$ and return to
  2.

\end{enumerate}

If this computation were done in a naive way, the number of stored
components would be $2^N$, thus rendering it unfeasible. Instead, the
wavefunctions may be stored as matrix product states (MPS):

\begin{equation}
\ket|\Psi>=\sum_{s_1\cdots s_N} \hbox{Tr}(A_1^{s_1} \cdots A_N^{s_N})
\ket|s_1,\cdots,s_N>
\label{mps}
\end{equation}

\noindent where the $A_i^{s_i}$ are $2N$ matrices of size $m\times m$,
and $m$ is called the {\em bond dimension}. The total number of
components in a MPS is, therefore, $2Nm^2$. Of course, $m$ must be
chosen {\em so that the ground state is always accurately
  represented}. The required value of the bond dimension $m$ will be,
therefore, of uttermost importance in order to evaluate the efficiency
of the procedure. 

Our technique of choice in order to determine the MPS representation
of the ground state for each value of $\Gamma$ is the density matrix
renormalization group (DMRG) \cite{White_93,Schollwock_11}. The
technique has several important features:

\begin{itemize}
\item DMRG is variational within the MPS subspace of the Hilbert space.
\item DMRG allows for adaptable values of $m$. In QWA simulations, $m$
  must be an {\em adaptable} parameter, which is chosen to be large enough
  to represent the state accurately at each simulation stage, to a
  given tolerance.
\item DMRG benefits from the use of a {\em seed state} in order to
  accelerate convergence, via the so-called {\em wavefunction
    transformations} \cite{White_96}.
\end{itemize}

Thus, the QWA algorithm under consideration is just an annealed
extension of the DMRG.

There are other algorithms to simulate a quantum computation based on
the MPS representation, e.g. the algorithm by Vidal \cite{Vidal_03}
for digital quantum computation or the one by Ba\~nuls and coworkers
\cite{Latorre_05} for AQC. The latter case uses real time simulation,
which may lead to a new source of loss of adiabaticity. Nonetheless,
they both point to similar relations between entanglement and the
efficiency of the calculation.


\section{Illustrating the Quantum Wavefunction Annealing}
\label{sec:illustration}

\begin{figure}
\epsfig{file=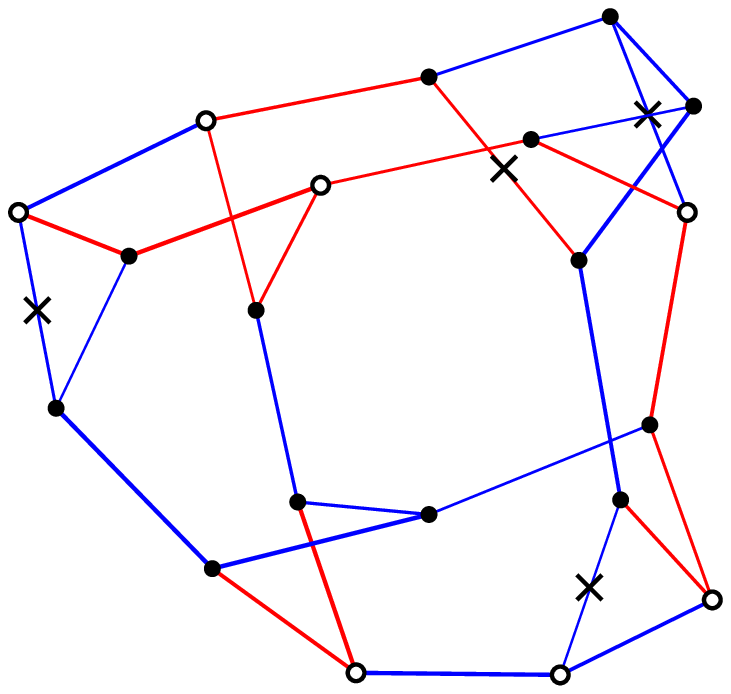,width=8cm}
\caption{\label{fig.graph}Illustration of a spin-glass problem
  instance solved by QWA. Red links are {\em antiferromagnetic}, and
  blue ones are {\em ferromagnetic}, with their width increasing with
  their strength. Sites are represented by either empty or full
  circles, denoting that the solution is either $+$ or $-$. The
  frustrated links in the optimal solution are marked by a cross.}
\end{figure}

The efficiency of QWA was analysed in \cite{Laguna_07} on the random
ITF model given by eq. (\ref{ritf}) for quasi-2D systems and random
graphs of fixed connectivity, and the results in that article give
support to the idea that the method will always reach the classical
minimal energy, provided that it is allowed to retain as many states
as desired. As a means of illustration, fig. (\ref{fig.graph}) shows a
sample graph with $N=20$ sites and connectivity $K=3$, along with the
QWA solution of the associated classical problem. Links are colored
either red (antiferromagnetic) or blue (ferromagnetic), with a width
proportional to its strength. Sites are depicted by either filled or
empty circles, representing $+$ and $-$ values of the spin. The links
marked by a cross are frustrated.

A technical point is in order: the final values of the spins are
obtained by measuring $S_z$ on each site. Of course, for the exact
GS of Hamiltonian (\ref{ritf}), those values will be zero. A
negligible longitudinal field in the $Z$-axis is introduced in order
to break the symmetry between the two identical solutions.

\begin{figure}
\epsfig{file=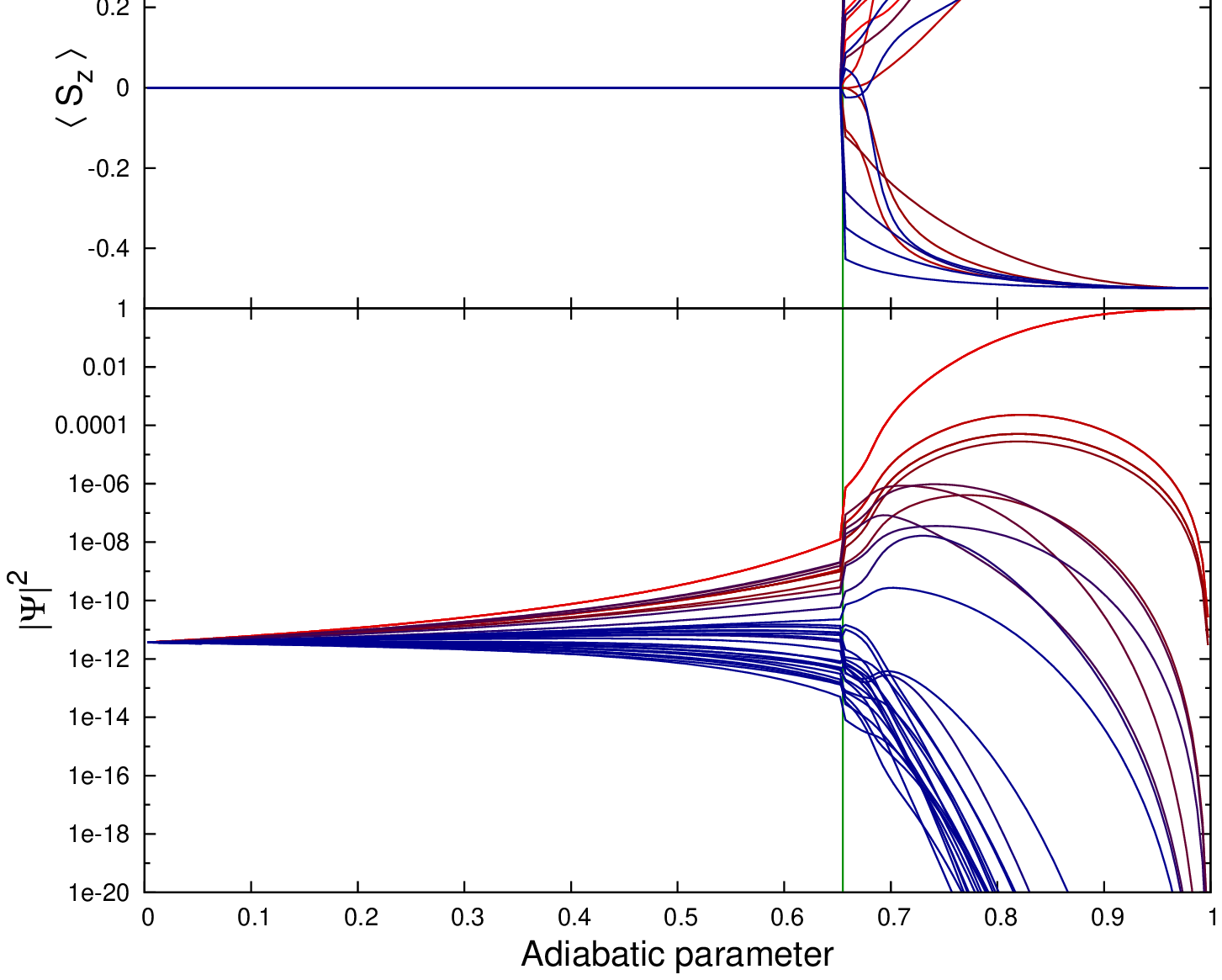,height=8.5cm,angle=270}
\caption{\label{fig.qwa}Development of the QWA algorithm. Top:
  evolution of the maximal von Neumann entanglement found at different
  stages of the simulation. Notice the peak at $\lambda_c\approx
  0.655$, denoting the quantum spin-glass phase transition. Center:
  Magnetization of each spin; notice the strong jump exactly at
  $\lambda_c$, and how not all the spins grow at the same
  rate. Bottom: Weights of different selected configurations,
  eq. (\ref{weight}), color denotes classical energy (red is
  lowest). They all start at $1/2^N$, and spread uniformly up to
  $\lambda_c$, when the transition takes place. Only the actual
  optimum reaches 1, all the rest fall eventually to zero.}
\end{figure}

Fig. (\ref{fig.qwa}) we illustrate how the QWA procedure develops. The
horizontal axis is common for all three plots, and represents the
advance of the adiabatic parameter $\lambda$. On the top panel, we
show the evolution of the maximal von Neumann entropy found in the
procedure. Notice the peak it presents at $\lambda_c\approx 0.655$,
which marks the quantum spin-glass phase transition. There are
secondary maxima, which are typical in those cases
\cite{Laguna_06}. The central panel shows the evolution of the
expectation values of $S_z$ on each spin during the QWA. For
$\lambda<\lambda_c$, i.e.: in the paramagnetic phase, the values are
very close to zero. At $\lambda_c$, they start a very fast increase,
whose rate is not uniform among them. For $\lambda=1$, they all take
values $+1/2$ or $-1/2$, and their signs will provide the solution of
the classical problem.
 
It is possible to define a {\em weight} for each classical spin
configuration $C=\{s_1\cdots s_N\}$:

\begin{equation}
|\Psi_C|^2 \equiv \left| \langle s_1\cdots s_N| \Psi \rangle \right|^2
\label{weight}
\end{equation}
i.e.: the probability for each configuration. Within the DMRG, it is
possible to obtain accurate estimates for the weights of different
configurations, just making use of the MPS structure
\cite{Laguna_07}. We have traced the weights for the actual optimal
configuration and a several others, and plotted them as a function of
$\lambda$ in the bottom panel of fig. (\ref{fig.qwa}). The color of
the line is an indicator of its energy as a classical configuration,
with red being the lowest energy and blue the highest. Notice that, up
to $\lambda_c$, all weights spread uniformly. At $\lambda_c$, the
quantum spin-glass phase transition takes place, and a few low-energy
configurations start to increase their weight, while most of them
decrease very fast. That increase, nonetheless, is only sustained by
the classical optimum, which reaches one, while all others fall to
zero eventually before $\lambda\to 1$.


\section{The DMRG-annealing conjecture}
\label{conjecture}

Let $M(N)$ represent the maximal bond dimension achieved during a
certain QWA procedure as a function of the system size. Let us now
state the {\em DMRG-annealing conjecture}:

\begin{enumerate}
\item The QWA algorithm, as described above, always obtains the true
  optimum.
\item The QWA-time is $O(N\cdot M(N)^k)$, for a certain $k$.
\end{enumerate}

In other words, QWA simulations will always obtain the true optimum of
the problem in a time that depends polynomially on the maximal bond
dimension. Thus, the QWA procedure will work in polynomial time in the
system size whenever the bond dimension grows at most polynomially
with $N$. Also, if the maximal bond dimension $M(N)$ does not grow
with $N$, the QWA time will scale linearly with the system size. In
this section we will discuss the arguments in favour of this
conjecture, and its possible pitfalls.

A first argument supporting the conjecture is the result by Vidal
\cite{Vidal_03} stating that a digital quantum computation which
involves a finite amount of entanglement can be efficiently simulated
using a classical computer. One should realize, nonetheless, that this
result, as such, is not applicable to the case of adiabatic quantum
computation. Indeed, digital and adiabatic quantum computation have
the same power \cite{Aharonov_04}, but their classical simulations
need not be equally amenable. The results of \cite{Vidal_03} have not
been extended to generic AQC on an arbitrary graph, but only to
dynamical simulations in 1D \cite{Vidal_04}.

Of course, the MPS representation of the wavefunction is highly
dependent on the {\em ordering} of the sites in the system, since the
maximal bond dimension, $M(N)$, can be strongly dependent on it. As an
example, the bond dimension for a non-critical 1D system saturates if
the natural ordering is chosen, but if our ordering is
$\{1,3,5,7,\cdots,2,4,6\cdots\}$, the bond dimension will grow
exponentially with the system size \cite{Poilblanc_10}. In QWA, the
ordering manifests in the choice of the {\em DMRG-path}. In absence of
a natural 1D structure, finding the path that minimizes the bond
dimension is a hard problem. It can be tackled approximately taking
profit of the {\em area law}, i.e.: assuming that entanglement (and,
thus, the bond dimension) of a division of the system will increase
with the number of broken bonds. Low-cost heuristical approaches to
this problem are discussed in \cite{Laguna_06,Laguna_07}.

Let us consider the eigenvalues of the reduced density matrix $\rho$
found at any stage of the DMRG procedure. They constitute a discrete
probability distribution, $\{p_i\}_{i=1}^{N_T}$, which we will assume
to be in decreasing order. Assuming a certain bond dimension $m$ is
equivalent to the approximation of retaining the first $m$ eigenvalues
$p_i$ and neglecting the rest. Given a certain tolerance $\epsilon>0$,
we would like to find $m(\epsilon)$, the minimum number of eigenvalues
that must be retained so that the sum of the remaining ones is smaller
than $\epsilon$:

\begin{equation}
\sum_{i> m(\epsilon)} p_i < \epsilon
\label{meff}
\end{equation}

The von Neumann entropy of entanglement is the Shannon entropy of the
eigenvalues of $\rho$: $S\equiv \< -\log(p_i) \>$. In the DMRG
literature it is usually assumed that the required bond dimension
$m(\epsilon)$ scales as the exponential of the von Neumann entropy of
entanglement \cite{Schollwock_11}, $m(\epsilon)\approx\exp(S)$. It is
straightforward to show that, given a MPS of dimension $m$, the
maximal von Neumann entanglement that it can sustain is, indeed,
$\log(m)$. Thus, $m(\epsilon)>\exp(S)$. But this scaling
$m(\epsilon)\approx \exp(S)$ can only be proved rigorously for some
mild distributions of the eigenvalues of the density matrices. If the
eigenvalues decay exponentially, it can be proved that for all
$\epsilon>0$, $m(\epsilon) \propto \epsilon^{-1} \exp(S)$. In case of
slower decays, still a polynomial relation can be found between
$\exp(S)$ and $m(\epsilon)$, but this is not true for generic
distributions.

Instead of the von Neumann entropy, we can employ a different measure
of entanglement, such as the average and variance of the {\em
  eigenvalue index}:

\begin{eqnarray}
\mu_i &\equiv& \sum_{i=1}^{N_T} i \cdot p_i \\
\sigma^2_i &\equiv& \sum_{i=1}^{N_T} i^2\cdot p_i - 
\mu_i^2 \equiv \<i^2\> - \<i\>^2
\label{variance.entropy}
\end{eqnarray}

With this new measure of entanglement, Chebyshev's inequality directly
provides the possibility of a rigorous relation:

\begin{equation}
m(\epsilon) =\mu_i+{1\over \sqrt{\epsilon}} \sigma_i
\quad\Rightarrow\quad \sum_{i>m(\epsilon)} p_i < \epsilon
\label{chebyshev}
\end{equation}

\begin{figure}
\epsfig{file=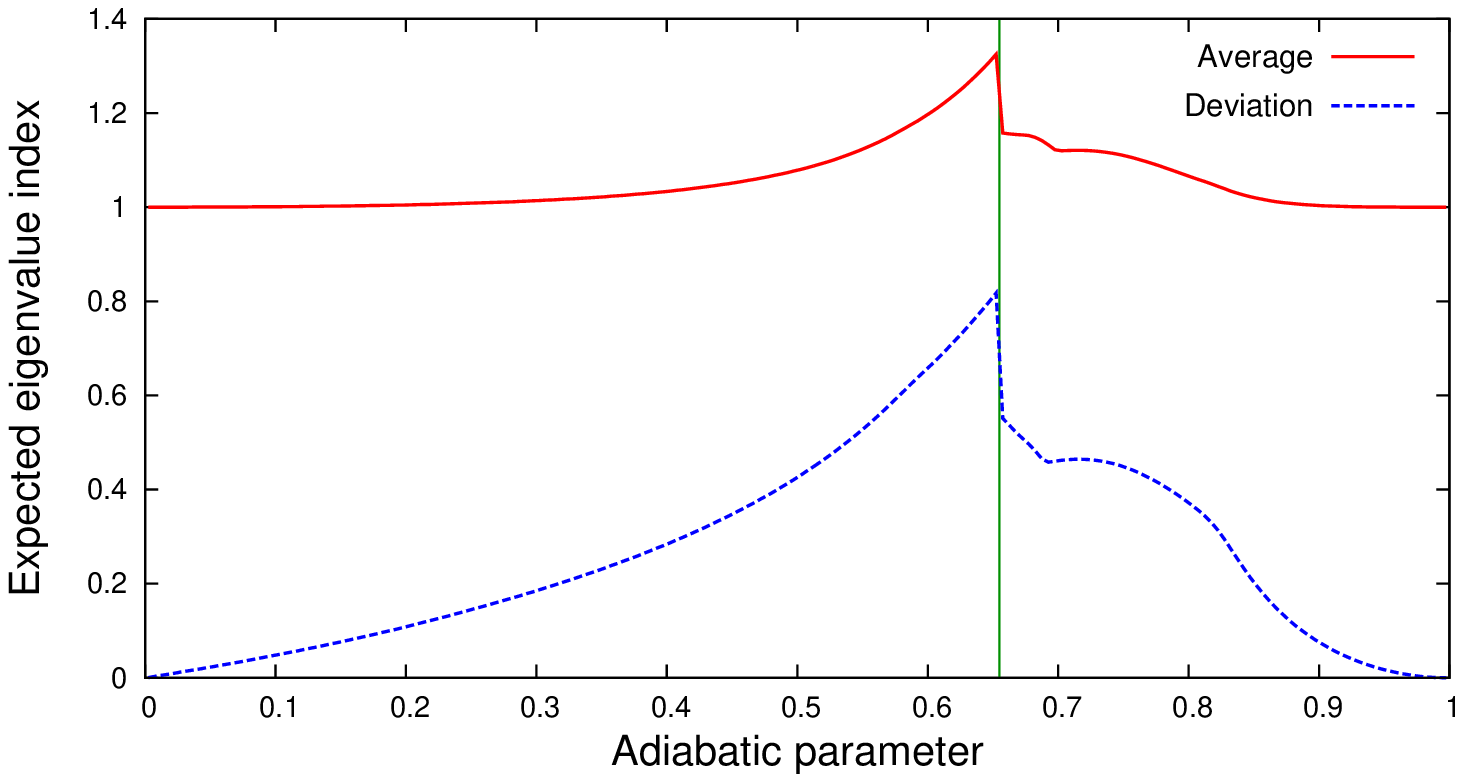,height=8.5cm,angle=270}
\caption{\label{fig.varent} Maximal average and maximal deviation of
  the eigenvalue index of the reduced density matrix during the QWA
  procedure described in sec. \ref{sec:illustration}.}
\end{figure}

Therefore, whenever $\mu_i$ and $\sigma_i$ are finite for every
partition during the DMRG process, we have a rigorous bound on the
bond dimension required for any given tolerance. Figure
(\ref{fig.varent}) shows the behavior of these measurements of
entanglement along the QWA process which served as illustration in
section (\ref{sec:illustration}). For this figure, we have obtained
the maximal value of $\mu_i$ and of $\sigma_i$, independently, as a
function of the adiabatic parameter.

Once that a technique has been provided to find the required bond
dimension $m(\epsilon)$ as a function of the DMRG tolerance, there is
a still a question to be addressed: {\em should the tolerance
  $\epsilon$ depend on the system size for the QWA algorithm to
  succeed with certainty?} The DMRG-annealing conjecture claims that
this is not the case \cite{Laguna_07}. The validity of this claim
rests only in limited empirical evidence, and requires further
investigation.

Another claim involved in the DMRG-conjecture is related to the speed
at which we are allowed to change $\lambda$. At each QWA step, the
algorithm attempts to transform the GS at a value $\lambda$ into the
GS at a slightly larger value, $\lambda+\delta\lambda$, by carrying
out as many DMRG sweeps as necessary. The cost of each DMRG sweep
scales polynomially in $m$. Thus, we need that the {\em number} of
DMRG sweeps does not scale appreciably with the system size, i.e.: the
states $\ket|\Psi(\lambda)>$ and $\ket|\Psi(\lambda+\delta\lambda)>$
must be so close that the transformation can be done in $O(1)$ DMRG
sweeps. This can always be achieved by reducing drastically
$\delta\lambda$, but this might increase drastically the computational
time. The claim is, therefore, that the {\em total} number of sweeps
required for the QWA computation, summing for all the values of
$\lambda$, does not scale appreciably with the system size. The
community consensus is that the DMRG is surprisingly robust in this
respect \cite{Schuch_08}, but no proof or counter-example of this
claim is known to us.

The QWA algorithm might be improved by taking an {\em adaptive} value
of $\delta\lambda$, estimated at each step from the fidelity,
$|\left<\Psi(\lambda)|\Psi(\lambda+\delta\lambda)\right>|$
\cite{Cozzini_06}. Finding a relation between these two values
ensuring a constant number of DMRG-sweeps would provide a valuable
insight into the previous claim, and also the means to accelerate the
procedure.

Some recent works have been devoted to study the efficiency of the
computation of MPS. Finding the ground state of a 1D quantum
Hamiltonian can be even {\bf NP}-complete
\cite{Eisert_06,Schuch_08}. If $m$ is fixed, instead of adaptable, MPS
are always nicely approximable \cite{Schuch_10}. This means that it is
always possible to obtain the best fixed-$m$ approximation to the GS
of a given Hamiltonian, within a factor $(1+\epsilon)$ of the ground
state energy, polynomially both in time and in $\epsilon^{-1}$. The
complexity class which conveys this is called {\em fully
  polynomial-time approximation scheme} ({\bf FPTAS})
\cite{Ausiello_99}.

Since the efficiency of the procedure is controlled by entanglement,
via the maximal-MPS dimension, let us summarize the known results
regarding its behaviour. Gapped systems usually fulfill the {\em area
  law} \cite{Sredniki_93,Plenio_10}, which states that the von Neumann
entanglement entropy $S$ between two parts of a system scales as the
number of broken links among them. This result was proved rigorously
in gapped 1D systems by Hastings \cite{Hastings_07}. In some higher
dimensional inhomogeneous and/or disordered systems it is known not to
hold \cite{Lewenstein_08,Rieger_07}. Moving to critical points, the
von Neumann entropy $S$ has proved to be a very useful magnitude in
order to pinpoint quantum phase transitions (QPT)
\cite{Laguna_06}. During any adiabatic process with finite $N$, $S$
always presents a local maximum at a QPT. It has also been conjectured
that a non-analyticity of $S$ may constitute a good indicator of a QPT
\cite{Chakravarti_08}. The 1D case is again rather special: at
criticality, $S$ will scale with the system size. In some cases, the
size-dependence of $S$ can be found via conformal field theory (CFT)
\cite{Vidal_02}. It has been shown that, for many critical 1D
problems, $S(N)\approx \alpha \log(N) + \beta$. Thus, assuming that
the bond dimension $m\approx\exp(S)$, it will grow polynomially with
the input size.


\section{\label{thesis}Physical implications of complexity theory}

\subsection{General principles}

Throughout this section, we will adopt the notation that a family of
Hamiltonians belongs to a complexity class if the problem of finding
the ground state of a generic instance in that family belongs to that
complexity class. Unless otherwise stated, we will assume the
DMRG-annealing conjecture to hold.

A complexity class which is simpler than {\bf P} is {\bf LIN}, i.e.:
the class of problems that can be solved in {\em linear} time. Let us
consider a {\em fixed} adiabatic route connecting Hamiltonians $H_0$
and $H_1$, such that $H_0$ belongs to a family in {\bf LIN} but $H_1$
belongs to a family of higher complexity. E.g., $H_1$ is taken from
{\bf P}, but with computational time scaling faster than $N$, perhaps
only $N\log(N)$. Then the bond dimension must {\em diverge} at some
point during the adiabatic route. Typically, this implies that the von
Neumann entropy will also diverge. This divergence is typically the
hallmark of a quantum phase transition (QPT).

The reason can be stated as follows. Let us assume that the
entanglement stays bounded during the whole AQC procedure. Now, let us
use a classical computer to run a QWA simulation of the AQC procedure,
in time $T(N)$. If the maximal bond dimension saturates with $N$, QWA
results asymptotically in a linear algorithm to obtain the ground
state of $H_1$, against the assumption. Therefore, entanglement must
grow without bound with the system size at some moment during the AQC
procedure, pointing to a QPT.

In the same line, if $H_0$ is {\bf P} and $H_1$ is {\bf NP}-complete,
and {\bf P}$\neq${\bf NP}, then any AQC connecting the two
Hamiltonians will find, at some moment, a state with maximal bond
dimension growing faster than polynomially in $N$. Typically this
implies that the von Neumann entropy will grow faster than
logarithmically with $N$. This state may correspond to a QPT. Again
the reason is easy to state: otherwise, the classical simulation will
provide a polynomial algorithm to solve an {\bf NP}-complete problem.

In general terms, we may say that the adiabatic connection of two
Hamiltonians within different complexity classes puts restrictions on
the physics along the path. In order to avoid violations of the
results from complexity theory, entanglement must diverge at some
moment during an AQC procedure. This divergence may take place as a
quantum phase transition of a certain kind. It can be regarded as a
kind of {\em quantum censorship} to prevent hard problems from being
solved easily.

Of course, the adiabatic route connecting $H_0$ and $H_1$ must be {\em
  fixed}, i.e.: established beforehand for all elements of the
family. If we allow it to be problem-dependent, there is always a
trivial way to find an adiabatic route with no entanglement which
requires to know the solution of the classical problem. E.g.: in order
to solve the spin-glass problem: (1) get the ground state of
$H_0=-\sum_i S^x_i$, (2) rotate each spin independently and
adiabatically until they reach their solution value $\sigma_i$, by
shifting to $H_{1/2}=-\sum_i \sigma_i S^z_i$ and (3) change
adiabatically to $H_1=-\sum_{\<i,j\>} J_{ij} S^z_i S^z_j$.

In this framework, it is easy to reconcile the apparently
contradictory results cited in the introduction. Vidal proved that a
digital quantum computation involving a finite amount of entanglement
can be efficiently simulated \cite{Vidal_03}, and Eisert proved that
the obtention of the GS of a Hamiltonian can be {\bf NP}-complete {\em
  even} if it is described by a MPS of low bond dimension
\cite{Eisert_06}. The DMRG-annealing conjecture predicts that any AQC
designed to find such GS, no matter the starting point or the
adiabatic route followed, will find a state with unbounded
entanglement, most probably a quantum phase transition. Thus, both
results are reconciled.

This analysis is independent of whether we focus on average or
worst-case complexity. Once the set of problems is characterized, and
a bound on the computational time is established, it can be
immediately converted into a bound for the entanglement entropy for an
AQC. 

A relevant point to be made is how to know whether the divergence of
the maximal entanglement points to a quantum phase transition or
not. In 1D it is known that at a QPT both the entanglement entropy and
the bond dimension diverge \cite{Hastings_07}. In higher dimensional
studies, a more careful analysis is required, since the maximal bond
dimension is likely to diverge along the whole AQC procedure. In any
case, for any finite instance of a problem, the maximum attained by
the entanglement determines the efficiency of the QWA. Therefore, if a
QPT is present, even if the bond dimension diverges with $N$ at all
points during the AQC, it is the scaling of entanglement at the QPT
which establishes the running time for the simulation.

\subsection{Concrete examples}

Let us return to the spin-glass hamiltonian (\ref{ritf}). When
$\lambda\to 0$, the obtention of the ground state is a trivial
problem, taking $O(1)$ time. In 1D, the obtention of the classical
spin-glass minimum energy state is obviously in {\bf LIN}. Therefore,
our results do not apply in this case, since QWA takes always time
$\geq N$.

In 2D, on the other hand, a prediction can be done. Solving the 2D
classical spin-glass problem is known to be in {\bf P}, but not in
{\bf LIN} \cite{Barahona_82}. Therefore, entanglement must diverge for
some value of $\lambda$. We can only state that the maximal entropy
will grow, {\em at least}, logarithmically. In fact, recent results
\cite{Rieger_07} (cleverly exploiting the properties of the infinite
randomness fixed point \cite{Fisher_95}, IRFP) show that it grows with
a modified area law: for a block division cutting $l$ links, the
entropy scales as $s(l)\approx l\log(\log(l))$. Maximal entropy, as it
is defined in this work, would be $S(N)\approx N^{1/2} \log(\log(N))$,
thus rendering the time for the QWA simulation exponential. Our result
is, therefore, valid but too weak.

Nonetheless, the previous expression for the block entropy in a 2D
quantum spin-glass is based on the average number of clusters cut by
the block division. A well designed DMRG path might never cut more
than one cluster at a time, just sweeping them one by one. In that
case, the maximal entropy might grow much more slowly with the system
size. But, in order to obtain such a path, one should {\em first}
solve the classical problem. Therefore, again, our basic result is not
violated.

In 3D, or for random graphs of fixed connectivity, the {\bf
  NP}-completeness of the problem forces the maximal entropy along the
route to grow faster than $\log(N)$. In this case, the result is not
surprising.

Other analysis have been carried out for the entanglement entropy
along typical standard quantum computations, and our general
statements also hold \cite{Latorre_04}. An AQC designed to solve the
{\em exact cover} problem (which is {\bf NP}-complete) found a QPT
with $S\approx N$ \cite{Latorre_04}. Although not an AQC, Shor's
algorithm also shows a similar behaviour. In this case, though, it is
not clear which is the complexity class of the problem under study
(i.e.: {\em integer factorization}) \cite{Latorre_04}. 

A different type of behavior is found in the {\em unsorted search}
problem, where the input is a set of $N$ values, among which we must
find a given one. This problem is in {\bf LIN} so, in consequence, an
AQC designed to solve using $N$ qubits may work with bounded
entanglement. Nonetheless, the standard adiabatic implementation of
Grover's algorithm \cite{Roland_02} works using only $n=\log_2(N)$
qubits, yet the maximal entanglement among them is also found to be
bounded \cite{Latorre_04}. This is apparently in contradiction with
our prediction, since DMRG would take a time polynomial in $n$, which
is always smaller than $N$, which is the classical computation
time. The explanation of this apparent paradox is that the approach in
\cite{Roland_02} makes use of an {\em oracle function}, which is a
non-local external element to the QWA formalism and can not be used in
the DMRG.

New predictions are easily made for AQC designed to solve problems
which have never been studied. Thus, an AQC designed to test planarity
of a graph, or 2-colorability, need not find a quantum phase
transition, since these problems belong to class {\bf LIN}. But an AQC
which sorts a set of numbers, or which performs the fast Fourier
transform, will find a divergence in the bond dimension and, very
likely, in the von Neumann entropy, since their running time is larger
than linear. The maximal entropy in those cases might grow very slowly
with size, since the (average) running time for the best algorithms
are $T\approx N\log(N)$, so our only bound is that $S$ should scale at
least like $\log(\log(N))$. On the other hand, if {\bf P}$\neq${\bf
  NP}, any AQC attempt to solve the traveling salesman problem, or
3-SAT, will always find maximal entropy growing faster than
logarithmically.


\section{\label{conclusions}Conclusions}

Theoretical physics has benefitted continuously from the incorporation
of the results of pure mathematics which were born without any
relation to it. Computational complexity theory is just another branch
of mathematics, and this work just attempts to extract its most
straightforward consequences for physics. As Nikolai Lobachevski put
it: {\em ``There is no branch of mathematics, however abstract, which
  may not someday be applied to the phenomena of the real world''}
\cite{Rose_88}.

In this work we have put forward a strategy to derive physical
inferences from computational complexity theory. If a physical process
is devised in order to solve some problem, simulating that process in
a (classical) computer constitutes a (classical) algorithm to obtain
the solution. The efficiency of this algorithm may be restricted by
complexity theory, and this restriction must have some counterpart in
the physical model which may apply to the real physical system.

Concretely, adiabatic quantum computation (AQC) may be simulated in
classical computers using quantum wavefunction annealing (QWA), which
is a simulation strategy based on the density matrix renormalization
group (DMRG). The efficiency of QWA is conjectured to be controlled by
the maximal entanglement attained during the physical
process. Different measures of entanglement are discussed
(bond dimension and von Neumann entropy) and a new one is introduced
(variance of the eigenvalue index). Arguments are given in favour of
this conjecture, along with an exposition of its possible pitfalls.

If there is a bound on the scaling of the computational time to solve
the problem on a classical computer, then this bound will transform
itself into another bound for the maximal entanglement attained during
the real physical procedure. This type of no-go reasoning bears
resemblance to the second law of thermodyamics. In this way, the
divergence of entanglement with the system size for some systems is
proved. This divergence, in some cases, may be viewed as the
apparition of a quantum phase transition, which can be regarded as a
``quantum censor'', preventing the solution of hard problems in an
easy way (in a classical computer).

The main possible pitfalls for the DMRG-annealing conjecture are the
following. It is not currently known what tolerance can be accepted in
the sum of neglected weights in DMRG in order to ensure the validity
of the QWA algorithm. The DMRG-annealing conjecture assumes that this
tolerance does not depend on the system size. Also, the total number
of DMRG sweeps needed to transform the wavefunction along the
adiabatic route is assumed to be independent of the system size. There
is limited empirical evidence supporting those claims. Our future work
will be devoted to the evaluation of the validity of these
assumptions.

The present derivation was performed using matrix product states (MPS)
and the DMRG, which are not specially well suited for multidimensional
systems, due to the need for a 1D path to run through the
system. Different generalizations of MPS exist, such as multiscale
entanglement renormalization Ansatz \cite{Vidal_08} (MERA) or
projected entangled pair states \cite{Schuch_07} (PEPS), which are
altogether labeled as tensor networks \cite{Nishino_01,Eisert_13}. New
techniques have been developed for 2D optimization problems, making
use of ideas related to {\em dynamic programming}
\cite{Hastings_08,Aharonov_10}. Also, other techniques have been
proposed in order to simulate real time evolution in the Heisenberg
picture \cite{Osborne_07}. We expect that application of this line of
thought to these sophisticated tools will provide stronger predictions
on the physics found during the performance of an adiabatic quantum
computation.


\begin{acknowledgments}
This work has been supported by the Spanish government by grants
FIS2012-33642 and FIS2012-38866-C05-1 and ERC grant QUAGATUA. The
authors wish to acknowlege very useful discussions with G. Sierra,
M. Lewenstein, I. Rodr\'{\i}guez, D. Peralta-Salas and E.H. Kim.
\end{acknowledgments}

%

\end{document}